\documentclass[nocopyright]{cimento}

\usepackage{graphicx}  
\usepackage{amsmath}	
\usepackage{amssymb}	
\usepackage[nosort]{cite}
\allowdisplaybreaks

\title{Evolution of an effective Hubble constant in $f(R)$ modified gravity}
\author{Tiziano Schiavone\from{ins:x}, 
Giovanni Montani\from{ins:y}\from{ins:z}

}

\instlist{\inst{ins:x} Galileo Galilei Institute for Theoretical Physics, Largo Enrico Fermi 2, I-50125 Firenze, Italy
\inst{ins:y} ENEA, Fusion and Nuclear Safety Department, C.R. Frascati, Via Enrico Fermi 45,  Frascati, I-00044 Rome, Italy
\inst{ins:z} Physics Department, ``Sapienza" University of Rome, Piazzale Aldo Moro 5, I-00185 Rome, Italy
  }

\begin{document}

\maketitle

\begin{abstract}

We investigate the Hubble constant tension within $f(R)$ modified gravity in the Jordan frame, focusing on its application to the dynamics of an isotropic Universe. A scalar field, non-minimally coupled to the metric, provides an extra degree of freedom compared to General Relativity. We analyze the impact of such a scalar field on the cosmic expansion, leading to an effective Hubble constant $H_{0}^{\text{eff}}(z)$, dependent on redshift $z$. We show that our $f(R)$ model might mimic dark energy and provide an apparent variation of the Hubble constant. Our results align with recent cosmological data analysis in redshift bins, indicating a decreasing trend of the Hubble constant. The redshift dependence of $H_{0}^{\text{eff}}(z)$ might potentially reconcile measurements of the Hubble constant from probes at different redshifts.   

\end{abstract}

\section{Introduction}\label{sec:intro}

The Hubble constant tension, one of the most significant issues in the era of `precision cosmology', is referred to the discrepancy between independent measurements of the Universe's current expansion rate, the Hubble constant $H_0$ \cite{DiValentino:2020zio,DiValentino:2021izs,Abdalla:2022yfr,Perivolaropoulos:2021jda,Vagnozzi:2023nrq}. The distance ladder method, based on direct observations and calibration of Cepheids and Type Ia supernovae (SNe Ia), provides a local measurement: $H_{0}^{\text{loc}}=73.04\pm1.04\,\textrm{km s}^{-1}\,\textrm{Mpc}^{-1}$ \cite{pantheon+}. On the other hand, the value inferred from the cosmic microwave background (CMB) radiation \cite{Planck:2018vyg} is $H_{0}^{\text{CMB}}=67.4\pm0.5,\textrm{km s}^{-1}\,\textrm{Mpc}^{-1}$, which is 4.9~$\sigma$ inconsistent with $H_{0}^{\text{loc}}$. The CMB estimate of $H_0$ relies on the $\Lambda$CDM cosmological model, which incorporates the cosmological constant $\Lambda$ and a cold dark matter component \cite{Weinberg:2008zzc}. 


The expansion rate for the fiducial flat $\Lambda$CDM cosmological model is expressed by the Hubble function $H^{\Lambda\textrm{CDM}}(z)$ in the Friedmann equation \cite{Weinberg:2008zzc}:
\begin{equation}
    H^{\Lambda\textrm{CDM}}(z)=H_0\,E^{\Lambda\textrm{CDM}}(z) \qquad \textrm{with} \quad E^{\Lambda\textrm{CDM}}(z) = \sqrt{\Omega_{m0}\left(1+z\right)^3+1-\Omega_{m0}}\,,
    \label{eq:E-LCDM}
\end{equation} 
where $\Omega_{m0}$ is the present cosmological density parameter of matter. Hence, $H_0$ is simply a constant by definition, being the expansion rate evaluated today, \textit{i.e.} $H(z=0)$. However, the $H_0$ tension challenges current cosmological models and may indicate systematic errors or new physics \cite{Vagnozzi:2019ezj,Hu:2023jqc}.

An interesting and original perspective to test cosmological models is to use an effective Hubble constant as a diagnostic tool in terms of the redshift $z$, as explained in \cite{Krishnan:2020vaf,Krishnan:2022}. Considering the Hubble function $H(z)$ and its reduced version $E(z)$ for a generic cosmological model, the ratio $H(z) / E(z)$ should be exactly the Hubble constant $H_0$. Here, $H(z)$ can be obtained from observations, while $E(z)$ is related to the theoretical model of the cosmic expansion, depending on the Universe content and the considered gravitational theory.
However, if data suggest a different cosmic evolution from that predicted theoretically from $E(z)$, then $H(z)$ and $E(z)$ are no more referred to the same model and a discrepancy occurs between them: the ratio $H(z) / E(z)$ is no longer constant. In other words, this ratio inevitably acquires a redshift dependence, resulting in an effective Hubble constant $H(z) / E(z) \equiv H_{0}^{\text{eff}}(z) \neq H_0 $, if we force data analysis to conform to the prescribed theoretical model that is not the correct one.

In this regard, possible evolving trends of $H_0$, $\Omega_{m0}$, or the marginalized absolute magnitude of SNe Ia have been recently discussed \cite{Wong:2019kwg,kazantzidis,Krishnan:2020obg,dainottiApJ-H0(z),DainottiGalaxies-H0(z),Schiavone:2022shz,Schiavone:2022wvq,Colgain:2022tql,2023arXiv230110572D,Jia:2022ycc,Montani:2023xpd,Malekjani:2023ple,colgain-bin,Liu:2024vlt}. In particular, two analysis in redshift bins \cite{dainottiApJ-H0(z),DainottiGalaxies-H0(z)} within the $\Lambda$CDM model pointed out an apparent variation of the Hubble constant as $H_{0}^{\text{fit}}(z)\sim(1+z)^{-\alpha}$ within 2~$\sigma$ confidence level, where $\alpha\sim 10^{-2}$. Additionally, extrapolating the fitting function $H_{0}^{\text{fit}}(z)$ up to the redshift of the last scattering surface, $z=1100$, local measurements of $H_0$ were reconciled with the CMB inferred value in 1~$\sigma$ \cite{dainottiApJ-H0(z)}.

Concerning theoretical attempts to accommodate the $H_0$ tension \cite{DiValentino:2021izs}, among various proposals, $f(R)$ modified gravity theories \cite{Nojiri-Odintsov2007,Sotiriou-Faraoni2010,Nojiri:2017ncd-nutshell} are considered promising scenarios due to an extra degree of freedom with respect to General Relativity (GR), given by a general function of the Ricci scalar $R$ in the gravitational Lagrangian, or equivalently a non-minimally coupled scalar field to the metric in the Jordan frame. The presence of this scalar field implies a variation of the gravitational coupling constant and affects the cosmological dynamics \cite{Schiavone:2022wvq,Montani:2023xpd,Odintsov:2020qzd,Nojiri:2022ski,Chen:2024wqc,Montani:2024xys}. For instance, $f(R)$ models might account for the present cosmic accelerated phase without the dark energy \cite{Hu:2007nk,Starobinsky:2007hu,Tsujikawa:2007xu}. However, in \cite{DainottiGalaxies-H0(z)}, it was shown that the Hu--Sawicki model \cite{Hu:2007nk}, one of the most robust $f(R)$ models in the late Universe, was inadequate to produce $H_{0}^{\text{eff}}(z)$. 

In this work, reporting the results obtained in \cite{Schiavone:2022wvq}, we present a $f(R)$ model able not only to mimic dark energy but also to provide the evolution of an effective Hubble constant $H_{0}^{\text{eff}}(z)$ with the redshift. In Sect.~\ref{sec:non-minimal}, we show that the dynamics of the scalar field in the Jordan frame can be interpreted as an evolving $H_{0}^{\text{eff}}(z)$ \cite{Schiavone:2022wvq}. In Sect.~\ref{sec:results}, we present the solution for the scalar field potential, associated with the function $f(R)$, to produce a decreasing trend for $H_{0}^{\text{eff}}(z)$. In Sect.~\ref{sec:conclusions}, we summarize our results.

\section{The non-minimally coupled scalar field and the cosmic expansion}\label{sec:non-minimal}

We show that $f(R)$ gravity implemented in a homogeneous and isotropic Universe might provide the evolution of an effective Hubble constant $H_{0}^{\text{eff}}(z)$ \cite{Schiavone:2022wvq}. The modified Friedmann equation in the equivalent scalar--tensor formalism in the Jordan frame of $f(R)$ gravity \cite{Nojiri-Odintsov2007,Sotiriou-Faraoni2010,Nojiri:2017ncd-nutshell} for pressure-less dust is written as:
\begin{equation}
    H^{2} =\frac{\chi\,\rho}{3\,\phi}+\frac{V\left(\phi\right)}{6\,\phi}-H\,\frac{\dot{\phi}}{\phi}\,,
    \label{eq:generalized-Friedmann}
\end{equation}
where $H(t)\equiv \frac{\dot{a}}{a}$ is the Hubble function, $a(t)$ is the scale factor, $\rho\left(t\right)$ is the energy density of the cosmological matter component, $\chi\equiv8\,\pi\,G$ is the Einstein constant, being $G$ the gravitational constant. Moreover, $\dot{}=d/dt$ and $t$ is the cosmic time. We recall that, in an isotropic Universe, all the relevant cosmological quantities depend only on $t$ and not spatial coordinates. We neglect relativistic components in the late Universe, since radiation is subdominant today. The extra degree of freedom in the Jordan frame compared to GR is represented by a scalar field $\phi\equiv df/dR$, which is non-minimally coupled to the metric, and whose dynamics is regulated by the potential $V\left(\phi\right)=\phi\,R\left(\phi\right)-f\left[R\left(\phi\right)\right]$. Note that an effective Einstein constant $\chi / \phi$ emerges from Eq.~\eqref{eq:generalized-Friedmann}.

Eq.~\eqref{eq:generalized-Friedmann} can be easily rewritten in terms of the redshift $z$ \cite{Schiavone:2022wvq} as 
\begin{equation}
    H(z) =\frac{H_0}{\sqrt{\phi(z)-\left(1+z\right)\,\phi^{\prime}(z)}}\sqrt{\Omega_{m0}\left(1+z\right)^3+\frac{V\left[\phi(z)\right]}{2\chi\rho_{c0}}}\,,
    \label{eq:general-Hubble-function}
\end{equation}
where $\phi^\prime\equiv d\phi/dz$. We used the definition $a_{0}/a=1+z$ with the present scale factor $a_{0}=1$, and the consequent relation $dz/dt=-\left(1+z\right)\,H\left(z\right)$. We recall that the matter component evolves like $\rho=\rho_{0}\left(1+z\right)^3$ with $\rho_0$ the present energy density of matter. The present critical energy density of the Universe is $\rho_{c0}\equiv3H_{0}^{2}/\chi$ and $\Omega_{m0}=\rho_{0}/\rho_{c0}$. Comparing Eq.~\eqref{eq:general-Hubble-function} with $H^{\Lambda\textrm{CDM}}(z)$ in the flat $\Lambda$CDM model, given by Eq.~\eqref{eq:E-LCDM}, it is clear that $V(\phi)$ might act as a varying cosmological constant.

Without any loss of generality, we split the potential profile into two parts: $V\left(\phi\right)\equiv 2\chi\rho_{\Lambda}+g\left(\phi\right)$, in which $\rho_{\Lambda}$ is the energy density associated with the cosmological constant in the flat $\Lambda$CDM model. To have a slight deviation from a cosmological constant picture, we assume that the function $|g\left(\phi\right)|\ll2\chi\rho_{\Lambda}$ for $0<z\lesssim z^*$, where $z^{*}\sim 0.3$ indicates the equivalence between dark energy and matter, such that $\Omega_{m0}\left(1+z^{*}\right)^{3}=\Omega_{\Lambda0}$. We recall that $\Omega_{\Lambda0}=\rho_{\Lambda}/\rho_{c0}=1-\Omega_{m0}$ in a flat Universe. Therefore, neglecting $g\left(\phi\right)$, Eq.~\eqref{eq:general-Hubble-function} can be rewritten as
\begin{equation}
    H\left(z\right)=H_{0}^{\text{eff}}(z)\,\sqrt{\Omega_{m0}\left(1+z\right)^3+1-\Omega_{m0}}\,,
    \label{eq:Hubble-parameter(z)}
\end{equation}
where the effective Hubble constant $H_{0}^{\text{eff}}(z)$ evolves with the redshift \cite{Schiavone:2022wvq} according to
\begin{equation}
    H_{0}^{\text{eff}}(z)\approx\frac{H_0}{\sqrt{\phi(z)-\left(1+z\right)\,\phi^{\prime}(z)}}\,.
    \label{eq:effective-H0}
\end{equation}
$H_0$ is really constant, while $H_{0}^{\text{eff}}(z)$ admits a redshift evolution due to the non-minimally coupled scalar field. In particular, to guarantee a decreasing trend of $H_{0}^{\text{eff}}$ with $z$, we require the condition $\phi\left(z\right)=\phi_0\,\left(1+z\right)^{2\alpha}$, where $\alpha>0$ is a constant and $\phi_0=\phi(z=0)$. Hence, $H_{0}^{\text{eff}}(z)\sim (1+z)^{-\alpha}$, according to the fitting function $H_{0}^{\text{fit}}(z)$ adopted in \cite{dainottiApJ-H0(z),DainottiGalaxies-H0(z)}.


Finally, we consider $H_{0}^{\text{eff}}(z)$ as a diagnostic tool \cite{Krishnan:2020vaf,Krishnan:2022}, following the discussion in Sect.~\ref{sec:intro}. Focusing on the mismatch between $H(z)$ in the Jordan frame of $f(R)$ gravity from Eq.~\eqref{eq:general-Hubble-function} and the $\Lambda$CDM function $E^{\Lambda\textrm{CDM}}(z)$ in Eq.~\eqref{eq:E-LCDM}, we end up in
\begin{equation}
    H_{0}^{\text{eff}}(z)=\frac{H(z)}{E^{\Lambda\textrm{CDM}}(z)} = \frac{H_0}{\sqrt{\phi(z)-\left(1+z\right)\,\phi^{\prime}(z)}}\sqrt{\frac{\Omega_{m0}\left(1+z\right)^3+\frac{V\left[\phi(z)\right]}{2\chi\rho_{c0}}}{\Omega_{m0}\left(1+z\right)^3+1-\Omega_{m0}}}\,,
\end{equation}
which is consistent with Eq.~\eqref{eq:effective-H0}, as we approximate the potential with its constant term.

\section{Analytical and numerical results}\label{sec:results}

We infer the profile of $V(\phi)$ from the full set of field equations in the Jordan frame, imposing Eq.~\eqref{eq:effective-H0} and the form of $\phi\left(z\right)=\phi_0\,\left(1+z\right)^{2\alpha}$. We stress that, in this calculation, we do not neglect the function $g(\phi)$, to check \textit{a posteriori} the approximation made for splitting $V(\phi)$ at low $z$. Combining the modified Friedmann equation, the modified acceleration equation, and the scalar field equation \cite{Nojiri-Odintsov2007,Sotiriou-Faraoni2010,Nojiri:2017ncd-nutshell}, after long but straightforward calculations (see \cite{Schiavone:2022wvq} for more details), we obtain a differential equation in $V(\phi)$, and we report here the analytical solution:
\begin{equation}
    \Tilde{V}\left(\phi\right)\, =\, \Tilde{V}\left(\phi_0\right)+\frac{6\alpha}{1-\alpha}\left\{\frac{2+\alpha}{\alpha}\frac{1-\Omega_{m0}}{\Omega_{m0}}\ln{\left(\frac{\phi}{\phi_0}\right)}+\frac{1+2\alpha}{3}\left[\left(\frac{\phi}{\phi_0}\right)^{\frac{3}{2\alpha}}-1\right]\right\}\,,
    \label{eq:soluzione-analitica-V(phi)}
\end{equation}
where $\Tilde{V}\equiv V/m^2$ is the dimensionless potential with $m^{2}\equiv \chi\rho_{0}/3=H_{0}^{2}\Omega_{m0}$. The constant $\Tilde{V}(\phi_0)=6\left(1-\Omega_{m0}\right)/\Omega_{m0}$ is set from the condition $V\left(\phi_0\right)= 2\chi\rho_{\Lambda}$. 

\begin{figure}
    \centering \includegraphics[scale=0.26]{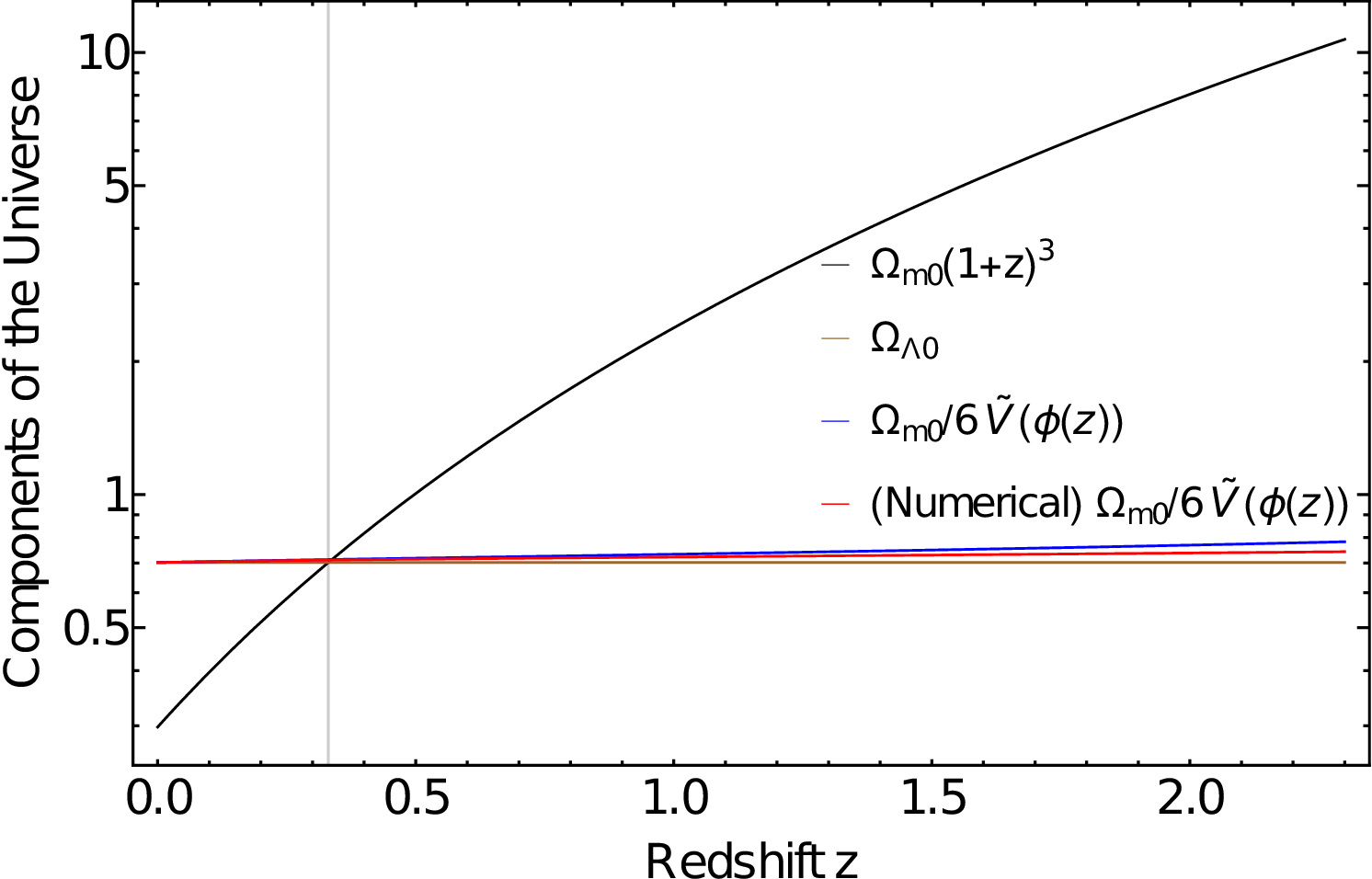}  \includegraphics[scale=0.26]{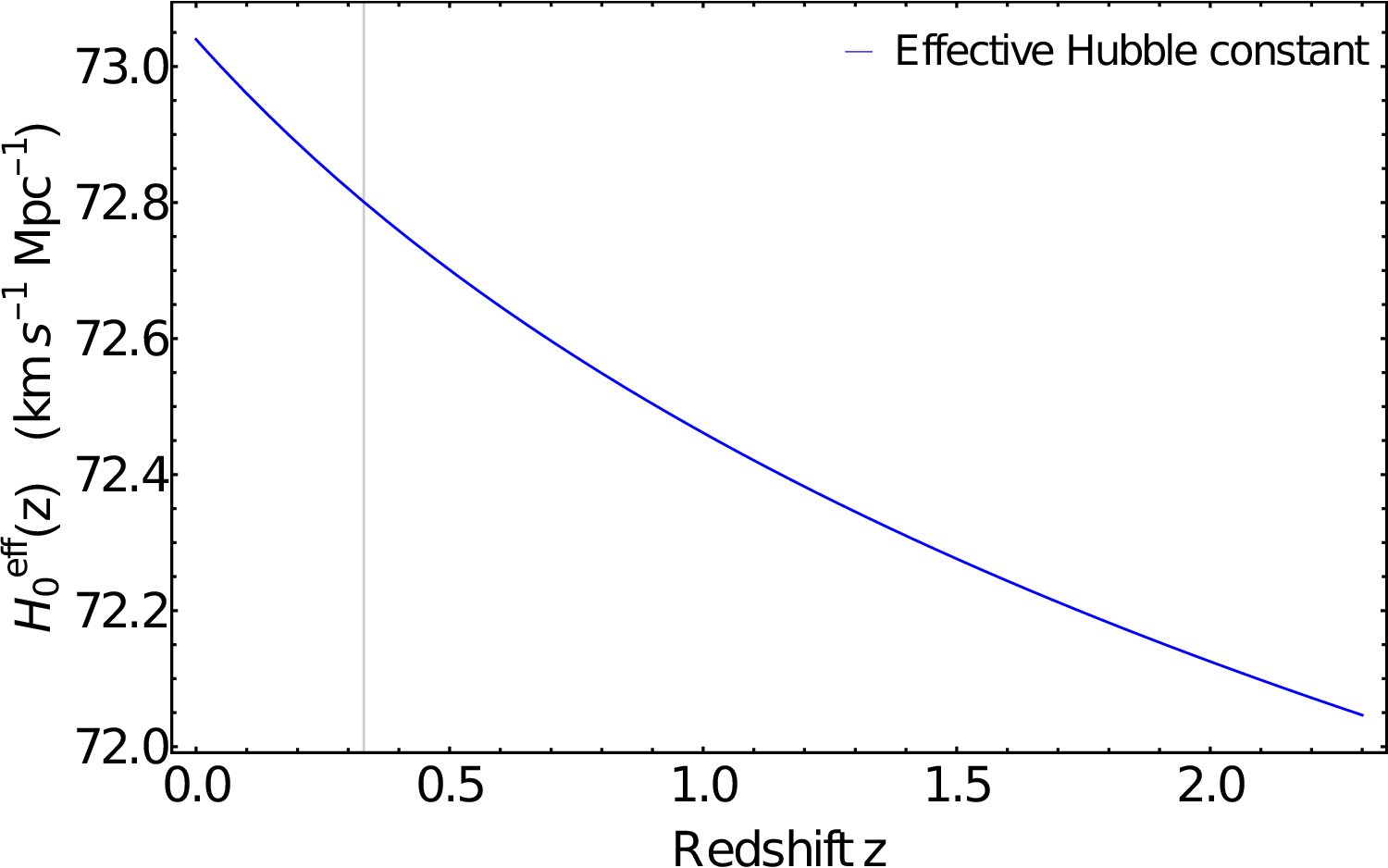}
    \caption{\textbf{Left panel}: evolution of the matter term (black line) with redshift $z$; contribution of the potential $\Tilde{V}\left(\phi\right)$ (blue line) able to produce $H_{0}^{\text{eff}}(z)$; the cosmological constant density parameter is indicated with a brown line. The numerical solution for $\Tilde{V}\left(\phi\right)$ is added with a red line \cite{Schiavone:2022wvq}. \textbf{Right panel}: decreasing trend of $H_{0}^{\text{eff}}$ with $z$. The grey vertical line denotes $z=z^*$.}
    \label{fig:scalar-field-potential}
\end{figure}

To quantify the slight deviation from GR ($\phi=1$), we fix $\phi_0=1-10^{-7}$ from solar system and cosmological constraints \cite{Hu:2007nk}. Combining Eq.~\eqref{eq:effective-H0} with $\phi\left(z\right)=\phi_0\,\left(1+z\right)^{2\alpha}$, and imposing that $H_0^{\textrm{eff}}\left(z=0\right)=H_{0}^\text{loc}$ \cite{pantheon+} and $H_{0}^{\text{eff}}(z=1100)=H_{0}^\text{CMB}$ \cite{Planck:2018vyg}, we obtain: $\alpha=1.1\times10^{-2}$, which is compatible in 1~$\sigma$ with the $\alpha$ used in \cite{dainottiApJ-H0(z)}, and $H_0=H_{0}^\text{loc} \sqrt{\phi_0 \left(1-2\alpha\right)}$. Once $H_0$ is fixed in Eq.~\eqref{eq:effective-H0}, the consequent relation for matter is $\Omega_{m0}=\chi \rho_0 / 3 H_0^2 =\Omega_{m0}^\text{loc}/\left[\phi_0 \left(1-2\alpha\right)\right]$, in which we set $\Omega_{m0}^\text{loc}=0.298$ \cite{scolnic-pantheon}. In Fig.~\ref{fig:scalar-field-potential}, we plot the different terms in the Hubble function \eqref{eq:general-Hubble-function} evolving with $z$, considering the matter component and the contribution related to the scalar field potential, \textit{i.e.} $\frac{V(\phi)}{2\chi\rho_{c0}}=\frac{\Tilde{V}\Omega_{m0}}{6}$, given by Eq.~\eqref{eq:soluzione-analitica-V(phi)}. For a comparison with the $\Lambda$CDM model, we also plot the cosmological constant contribution $\Omega_{\Lambda0}$. 

To check our initial assumption on a dominant constant term in the potential for $0<z\lesssim z^*$, we note from Fig.~\ref{fig:scalar-field-potential} that $\frac{\Tilde{V}\Omega_{m0}}{6}$ coincides with $\Omega_{\Lambda0}$ for $z=0$, while we estimate a relative difference of $1.6\%$ at $z=z^*$. Hence, $\frac{\Tilde{V}\Omega_{m0}}{6}$ can be regarded as a slowly varying cosmological constant, which is also responsible for $H_{0}^{\text{eff}}(z)$.

Additionally, we solve numerically the full set of field equations in the Jordan frame of $f(R)$ gravity without any assumption on the potential profile at low $z$ but only requiring the presence of $H_{0}^{\text{eff}}(z)$. We observe from Fig.~\ref{fig:scalar-field-potential} that analytical and numerical solutions of $\Tilde{V}\left(\phi\right)$ mostly overlap in the range corresponding to $0<z\lesssim z^*$. 

The right panel in Fig.~\ref{fig:scalar-field-potential} shows the decreasing trend of the effective Hubble constant $H_{0}^{\text{eff}}$ with $z$. Therefore, $H_{0}^{\text{eff}}(z)$ might potentially solve the $H_0$ tension, matching incompatible measurements of $H_{0}^{\text{loc}}$ and $H_{0}^{\text{CMB}}$ \cite{dainottiApJ-H0(z),DainottiGalaxies-H0(z),Schiavone:2022wvq} at different redshifts.  

Finally, we conclude this section by mentioning that, in \cite{Schiavone:2022wvq}, we compute analytically the functional form $f(R)$ associated with the potential $V\left(\phi\right)$ in the low-redshift regime. In this regard, we expand the solution $\phi(z)$ and the potential $\Tilde{V}\left(\phi\right)$ in Eq.~\eqref{eq:soluzione-analitica-V(phi)} for $z\ll1$ (or equivalently $\phi$ around $\phi_0$) up to the second order. Then, using the field equation $R=dV/d\phi$ and the relation $f\left(R\right)=R\,\phi\left(R\right)-V\left[\phi\left(R\right)\right]$ in the Jordan frame of $f(R)$ gravity, we obtain constant, linear, and quadratic corrections in $R$ and $R^2$ as in the quadratic $f(R)$ gravity \cite{STAROBINSKYquadratic,fanizza-f(R),fanizza-quadratic}. See \cite{Schiavone:2022wvq} for explicit calculations. Note that the analytical expression of $f(R)$ obtained in \cite{Schiavone:2022wvq} is only an approximated solution valid for $z\ll1$. Moreover, the $\Lambda$CDM model is recovered for $\alpha\rightarrow0$ and $\phi_0\rightarrow1$.


\section{Conclusions}\label{sec:conclusions}

We studied the $H_0$ tension in $f(R)$ modified gravity in the Jordan frame, focusing on how the non-minimally coupled scalar field $\phi$ affects the expansion of an isotropic Universe. We showed that a varying effective Hubble constant $H_{0}^{\text{eff}}(z)$ emerges (see Eq.~\eqref{eq:effective-H0}), even though we set the same matter content in the $\Lambda$CDM model.  

We determined the profile of the potential $V\left(\phi\right)$ \cite{Schiavone:2022wvq}, which regulates the dynamics of $\phi$ and allows the presence of $H_{0}^{\text{eff}}(z)$. The potential represents a slowly varying cosmological constant (see Fig.~\ref{fig:scalar-field-potential}), extending the $\Lambda$CDM paradigm. The resulting $f(R)$ functional form is computed for $z\ll1$ \cite{Schiavone:2022wvq}, pointing out a quadratic $f(R)$ model.

Our study successfully addresses two critical aspects simultaneously: firstly, it proposes a $f(R)$ model to provide the cosmic acceleration in the late Universe; secondly, it introduces an evolving Hubble constant $H_{0}^{\text{eff}}(z)$. The possibility to theoretically predict the power-law behavior of $H_{0}^{\text{eff}}(z)$ offers an intriguing perspective to validate models via suitable data analysis toward an effective Hubble constant. 

If the variation of $H_{0}^{\text{eff}}(z)$ is sufficiently rapid, we can expect to appreciate the rescaling of $H_{0}^{\text{eff}}(z)$ within the same population of sources. In particular, this scenario offers a natural explanation for the fitting function $H_0^{\text{fit}}(z)$, observed in \cite{dainottiApJ-H0(z),DainottiGalaxies-H0(z)} via a binned analysis of the Pantheon sample of SNe Ia \cite{scolnic-pantheon}. The emergence of a varying effective Hubble constant from cosmological data analysis is a significant proof that $H_{0}^{\text{eff}}(z)$ has a precise physical meaning, related to real measured values of the Hubble constant at the redshift of a given source. These results and the future increasing quality of statistical analysis of SNe Ia should foster the SH0ES collaboration to search for a similar feature in their upgraded samples, overcoming the binning criticism in \cite{Brout:2020bbg}.

We stress that, while the power-law behavior of $H_{0}^{\text{eff}}(z)$ is a regular smooth decaying, other models \cite{Montani:2023xpd,Montani:2023ywn} provide an effective Hubble constant that reaches rapidly $H_{0}^{\text{CMB}}$ for $z\gtrsim5$. In this regard, in the future, the combination of SNe Ia data with other farther sources, like quasars and gamma-ray bursts, will be crucial to better characterize the evolution of $H_{0}^{\text{eff}}(z)$ and shed new light on the $H_0$ tension \cite{Lenart:2022nip,Bargiacchi:2023jse,Dainotti:2023bwq}.

This study highlights the need for further investigations, as the redshift approaches the CMB observations, to determine if $H_{0}^{\text{eff}}(z)$ can adequately solve the Hubble tension.


\acknowledgments
The work of TS is supported by the Galileo Galilei Institute with the Boost Fellowship. 



\begin{thebibliography}{100}

\bibitem{DiValentino:2020zio} \BY{Di Valentino E. \atque others}
\IN{Astropart. Phys.}{131}{2021}{102605}.  

\bibitem{DiValentino:2021izs} \BY{Di Valentino E., Mena O. \atque others}
\IN{Class. Quant. Grav.}{38}{2021}{153001}.

\bibitem{Abdalla:2022yfr} \BY{Abdalla E., Abellán G.~F., Aboubrahim A. \atque others}
\IN{JHEAp}{34}{2022}{49}.

\bibitem{Perivolaropoulos:2021jda} \BY{Perivolaropoulos L. \atque Skara F.}
\IN{New Astron. Rev.}{95}{2022}{101659}.

\bibitem{Vagnozzi:2023nrq} \BY{Vagnozzi S.}
\IN{Universe}{9}{2023}{393}.

\bibitem{pantheon+} \BY{Riess A.~G., Yuan W., Macri L.~M. \atque others}
\IN{ApJ}{934}{2022}{L7}.
 
\bibitem{Planck:2018vyg} \BY{Aghanim N., Akrami Y., Ashdown M. \atque others}
\IN{A\&A}{641}{2020}{A6}.

\bibitem{Weinberg:2008zzc} \BY{Weinberg S.}
\TITLE{Cosmology}, (Oxford Univ. Press, Oxford) 2008.

\bibitem{Vagnozzi:2019ezj} \BY{Vagnozzi S.}
\IN{Phys. Rev. D}{102}{2020}{023518}.

\bibitem{Hu:2023jqc} \BY{Hu J.-P. \atque Wang F.-Y.}
\IN{Universe}{9}{2023}{94}.

\bibitem{Krishnan:2020vaf} \BY{Krishnan C., Colg\'ain E.~\'O. \atque others}
\IN{Phys. Rev. D}{103}{2021}{103509}.

\bibitem{Krishnan:2022} \BY{Krishnan C. \atque Mondol R.}
preprint arXiv:2201.13384.

\bibitem{Wong:2019kwg} \BY{Wong K.~C., Suyu S.~H., Chen G.~C.-F. \atque others}
\IN{MNRAS}{498}{2020}{1420}.

\bibitem{kazantzidis} \BY{Kazantzidis L. \atque Perivolaropoulos L.}
\IN{Phys. Rev. D}{102}{2020}{023520}.

\bibitem{Krishnan:2020obg} \BY{Krishnan C., Colg\'ain E.~\'O. \atque others}
\IN{Phys. Rev. D}{102}{2020}{103525}.

\bibitem{dainottiApJ-H0(z)} \BY{Dainotti M. G., De Simone B., Schiavone T. \atque others}
\IN{ApJ}{912}{2021}{150}.

\bibitem{DainottiGalaxies-H0(z)} \BY{Dainotti M.~G., De Simone B., Schiavone T. \atque others}
\IN{Galaxies}{10}{2022}{24}.

\bibitem{Schiavone:2022shz} \BY{Schiavone T., Montani G., Dainotti M.~G. \atque others}
preprint arXiv:2205.07033.

\bibitem{Schiavone:2022wvq} \BY{Schiavone T., Montani G. \atque Bombacigno F.}
\IN{MNRAS Letters}{522}{2023}{L72}.

\bibitem{Colgain:2022tql} \BY{Colg{\'a}in E.~\'O., Sheikh-Jabbari M.~M. \atque Solomon R.}
\IN{Phys. Dark Univ.}{40}{2023}{101216}

\bibitem{2023arXiv230110572D} \BY{Dainotti M.~G., De Simone B. \atque others}
\IN{PoS}{CORFU2022}{2023}{235}

\bibitem{Jia:2022ycc} \BY{Jia X.~D., Hu J.~P. \atque Wang F.~Y.}
\IN{A\&A}{674}{2023}{A45}.

\bibitem{Montani:2023xpd} \BY{Montani G., De Angelis M., Bombacigno F. \atque Carlevaro N.}
\IN{MNRAS Letters}{527}{2023}{L156}.

\bibitem{Malekjani:2023ple} \BY{Malekjani M., Mc Conville R., Colg\'ain E.~\'O. \atque others}
\IN{EPJC}{84}{2024}{317}.

\bibitem{colgain-bin} \BY{Colg{\'a}in E.~\'O., Sheikh-Jabbari M.~M., Solomon R. \atque others}
\IN{Phys. Dark Univ.}{44}{2024}{101464}

\bibitem{Liu:2024vlt} \BY{Liu Y., Yu H. \atque Wu P.}
\IN{Phys. Rev. D}{110}{2024}{L021304}.

\bibitem{Nojiri-Odintsov2007} \BY{Nojiri S. \atque Odintsov S. D.}
\IN{eConf}{C0602061}{2006}{06}.

\bibitem{Sotiriou-Faraoni2010} \BY{Sotiriou T.~P. \atque Faraoni V.}
\IN{Rev. Mod. Phys.}{82}{2010}{451}.

\bibitem{Nojiri:2017ncd-nutshell} \BY{Nojiri S., Odintsov S.~D. \atque Oikonomou V.~K.}
\IN{Phys. Rept.}{692}{2017}{1}.

\bibitem{Odintsov:2020qzd} \BY{Odintsov S.~D., S\'aez-Chill\'on G\'omez D. \atque Sharov G.~S.}
\IN{Nucl. Phys. B}{966}{2021}{115377}.

\bibitem{Nojiri:2022ski} \BY{Nojiri S., Odintsov S.~D. \atque Oikonomou V.~K.}
\IN{Nucl. Phys. B}{980}{2022}{115850}.

\bibitem{Chen:2024wqc} \BY{Chen H., Katsuragawa T., Nojiri S. \atque Qiu T.}
preprint arXiv:2406.16503.

\bibitem{Montani:2024xys} \BY{Montani G., Carlevaro N. \atque De Angelis M.}
preprint arXiv:2407.12409.

\bibitem{Hu:2007nk} \BY{Hu W. \atque Sawicki I.}
\IN{Phys. Rev. D}{76}{2007}{064004}.

\bibitem{scolnic-pantheon} \BY{Scolnic D.~M., Jones D.~O., Rest A. \atque others}
\IN{ApJ}{859}{2018}{101}.

\bibitem{Starobinsky:2007hu} \BY{Starobinsky A.~A.}
\IN{JETP Lett.}{86}{2007}{157--163}.

\bibitem{Tsujikawa:2007xu} \BY{Tsujikawa S.}
\IN{Phys. Rev. D}{77}{2008}{023507}.

\bibitem{STAROBINSKYquadratic} \BY{Starobinsky A.~A.}
\IN{Phys. Lett. B}{91}{1980}{99}.

\bibitem{fanizza-f(R)} \BY{Cosmai L., Fanizza G. \atque Tedesco L.}
\IN{Int. J. Theor. Phys.}{55}{2016}{754}.

\bibitem{fanizza-quadratic} \BY{Fanizza G., Franchini G., Gasperini M. \atque Tedesco L.}
\IN{Gen. Rel. Grav.}{52}{2020}{111}.

\bibitem{Brout:2020bbg} \BY{Brout D., Hinton S. \atque Scolnic D.}
\IN{ApJ Lett.}{912}{2021}{L26}

\bibitem{Montani:2023ywn} \BY{Montani G., Carlevaro N. \atque Dainotti M.~G.}
\IN{Phys. Dark Univ.}{44}{2024}{101486}

\bibitem{Lenart:2022nip} \BY{Lenart A.~L., Bargiacchi G., Dainotti M. \atque others}
\IN{ApJ Suppl.}{264}{2023}{46}

\bibitem{Bargiacchi:2023jse} \BY{Bargiacchi G., Dainotti M.~G., Nagataki S. \atque Capozziello S.}
\IN{MNRAS}{521}{2023}{3909}

\bibitem{Dainotti:2023bwq} \BY{Dainotti M.~G., Bargiacchi G., Bogdan M. \atque others}
\IN{ApJ}{951}{2023}{63}


\end{thebibliography}
\end{document}